# All-PM Divided Pulse Fiber Oscillator Mode-locked with the Optical Kerr-effect

**Marvin Edelmann,**[1,2,3] **Yi Hua,**[1,4] **Gabor Kulcsar**[5] **and Franz X. Kärtner**[1,4,*]

[1] *Center for Free-Electron Laser Science (CFEL), DESY, Notkestr. 85, 22607 Hamburg, Germany*
[2] *Department of Physics, Universität Oldenburg, Ammerländer Heerstr. 114-118, 26111 Oldenburg, Germany*
[3] *Cycle GmbH, Notkestr. 85, 22607 Hamburg, Germany*
[4] *Department of Physics, Universität Hamburg, Jungiusstr. 9, 20355 Hamburg, Germany*
[5] *Laser Impulse, Dr. Gabor Kulcsar, Wiesenkamp 30, 24226 Heikendorf, Germany*
*\*Corresponding author: franz.kaertner@desy.de*

**In this letter, we investigate a Yb-doped mode-locked fiber oscillator that uses coherent pulse division and recombination to avoid excessive nonlinear phase shifts. The mode-locking mechanism of the laser is based on the accumulation of a differential nonlinear phase between orthogonal polarization modes in the polarization-maintaining fiber segment. The inserted coherent pulse divider, based on YVO$_4$-crystals rotated successively by 45°, enables stable and undistorted mode-locked steady-states. The output pulse energy is increased from 89 pJ in the non-divided operation by ~6.5 dB to more than 400 pJ with three divisions. Measurements of the amplitude-fluctuations reveal a simultaneous broadband reduction of up to ~9 dB in the frequency range from 10 kHz to 2MHz.**

The domain of applications for mode-locked fiber oscillators is constantly growing and already includes many scientific fields and state-of-the-art technologies such as frequency metrology [1], seeding high-power amplifier chains [2], synchronization and timing distribution [3] and optical microwave generation [4]. The rapid progress in the above-mentioned applications necessitates a constant improvement of the fiber oscillator these systems are based on in terms of the environmental stability and output characteristics such as the obtainable pulse energy and noise performance. A key component of ultrafast fiber oscillators is the saturable absorber (SA) which often determines the cavity structure and the achievable output parameters [5-7]. Among the variety of mode-locking mechanisms, *artificial* SAs based on the optical Kerr-effect revealed themselves as promising candidates for many applications due to their environmentally stable cavity structures based on polarization-maintaining fibers [8], self-starting operation [9] and superior noise performance [10-12]. A well-established technique utilizes asymmetric fiber loops known as nonlinear amplifying/optical loop mirrors (NALM/NOLM), often in combination with a non-reciprocal phase bias that is required for self-starting operation [13]. In these variations of a Sagnac-interferometer, the self-amplitude modulation has its origin in the accumulation of a nonlinear phase difference between counter-propagating pulses in the fiber loop [14]. Fermann et. al. proposed an alternative linear cavity structure that allows Kerr-type mode-locking instead by using co-propagating orthogonal polarization modes in a PM-fiber to accumulate the required nonlinear phase difference with a compensation of linear phase shifts [15]. Despite the advantages of the mentioned oscillator structures in all-PM configuration mode-locked with the optical Kerr-effect, their performance is still fundamentally limited by the roundtrip nonlinear phase shift and the occurrence of multiple pulse formation due to soliton splitting or wave-breaking if the pulse energy exceeds a certain threshold [16,17]. The achievable intracavity power further restricts the pulse energy and the laser noise performance in terms of phase noise, timing-jitter and amplitude-fluctuations at the output [18-20]. Different approaches have been proposed to reduce the roundtrip nonlinear phase shift. Besides scaling the fiber core size with the implementation of large-mode area (LMA) fibers [21], it is possible to introduce a large breathing ratio of the pulse duration and thus reduce the nonlinear phase shift per roundtrip by accessing the dissipative or dispersion-managed soliton regime through an engineered all-normal dispersion (ANDi) or a net-dispersion close to zero, respectively [22,23]. In 2014, Wise et. al. further proposed the application of a coherent divided pulse amplification (DPA) scheme to reduce the pulse peak power in a non-PM, SESAM mode-locked oscillator leading to a significant increase of the output pulse energy [24]. While the concept of DPA is routinely used in amplifier systems nowadays [25,26], there is not yet any information available on the possibility of a stable implementation within an all-PM oscillator cavity with artificial SA. Consequently, in this letter we demonstrate and investigate the first time the implementation of a coherent division and recombination scheme in such an all-PM cavity structure to increase the obtainable pulse energy and the noise performance. The Kerr-type SA mechanism is based on the accumulation of a differential nonlinear phase shift $\Delta\varphi_{nl}$ between orthogonal polarization modes in the PM-fiber segment as proposed in Ref. [15]. The resulting sinusoidal transmission function $T(\Delta\varphi_{nl})$ in combination with a non-reciprocal phase bias allows for stable and undistorted mode-locking. The newly implemented pulse divider scheme consists of multiple number of YVO$_4$- crystals rotated by 45° and the birefringent fiber segment itself. In the experiment, the achievable output power is increased by up to 6.5 dB with 3 pulse divisions compared to the fundamental mode-locked state with no pulse divisions. Measurements of the RF-spectrum and the intensity-fluctuations reveal a reduced noise-level in the offset-frequency range from 10 kHz to ~2 MHz for an increasing number of divisions.

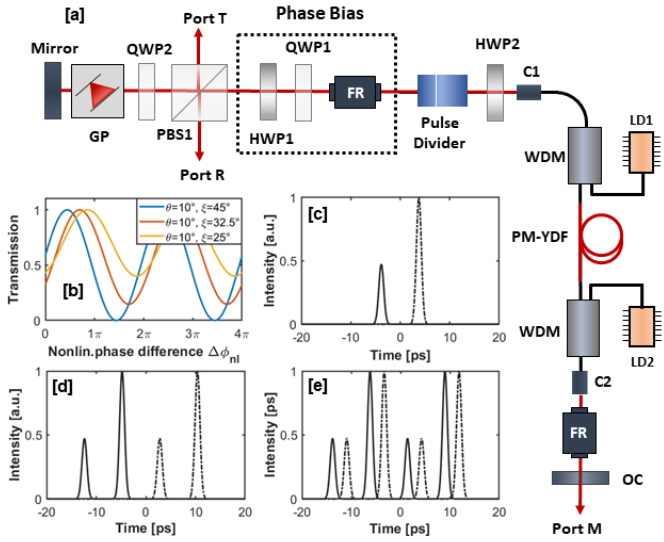

**Fig.1:** [a]: Experimental setup of the all-PM divided pulse oscillator. GP: Grating pair, QWP: Quarter-wave plate, PBS: Polarizing beam-splitter, HWP: Half-wave plate, FR: Faraday-rotator, LD: Laser diode, WDM: Wavelength division multiplexer, YDF: Ytterbium-doped fiber, OC: Output coupler, C: fiber collimator. [b]: Transmission function T($\Delta\varphi_{nl}$) for different rotation angles of HWP1 ($\theta$) and QWP1 ($\xi$). [c]: Simulated divided pulse train behind the 10 mm YVO$_4$-crystal (45°-rotation to the transmission axis of PBS1) with one pulse parallel to the slow (solid) and fast axis (dashed) for $\theta = 10°$ and $\xi = 25°$. [d]: Divided pulses after an additional, 45°-rotated, 20 mm YVO$_4$-crystal in the divider. [e]: Divided pulses at Port M with third division from the PM-fiber segment.

A schematic of the all-PM divided pulse oscillator (DPO) is shown in Fig.1 [a]. The 2.2 m fiber segment contains a 0.6 m highly Yb-doped gain fiber (CorActive Yb-401 PM) that is bi-directionally pumped by two 0.9 W laser diodes at 976 nm (Thorlabs BL976-PAG900), coupled into the fiber with wavelength-division multiplexers (WDM). In the free-space arm on the C2 collimator side, the light is transmitted through a 45°-Faraday-rotator (FR, single-pass) and is subsequently partially reflected back in the cavity by an output coupler (OC, 15% transmission) that enables the analysis of the divided pulse bursts through the monitor Port M. The fiber output at collimator C1 is followed by the second free-space arm containing a YVO4-based pulse divider and the non-reciprocal phase bias that consists of a 45°-FR, a quarter-wave plate (QWP1) and a half-wave plate (HWP1). A 1000 lines/mm polarization-insensitive transmission grating pair allows tunable dispersion-management. The repetition rate of the laser is 36.7 MHz and the cavity net-dispersion is fixed at ~-104 · 10$^{-3}$ ps$^2$ which ensures a mode-locked operation in the soliton regime. The soliton regime is chosen for the experiment to ensure a short pulse duration of <1 ps to avoid a time-wise overlap of the divided pulses in the fiber segment. In the fundamental state of the oscillator without YVO$_4$-crystals, the orthogonal polarization modes in the PM-fiber segment accumulate a nonlinear phase difference $\Delta\varphi_{nl}$ provided that there is an asymmetric energy splitting ratio ε [27]. In this case, the value of ε together with the shape of the sinusoidal SA transmission function T($\Delta\varphi_{nl}$) is fully determined by the settings of the waveplate angles $\theta$ (HWP1) and $\xi$ (QWP1) in the phase bias as derived e.g., in Ref. [10]. Besides an accumulation of $\Delta\varphi_{nl}$, the PM fiber's birefringence of B=0.381 ·10$^{-3}$ (Coherent PM980-XP) further results in a drift-off between the polarization modes with ~1.3 ps/m that is compensated through the Faraday-mirror that ensures an identical optical path for both polarization modes through the 90°-rotated back-reflection. This unique cavity structure enables the application of the YVO$_4$-based DPA scheme as the linear phase shift compensation ensures the coherent recombination of divided pulses.

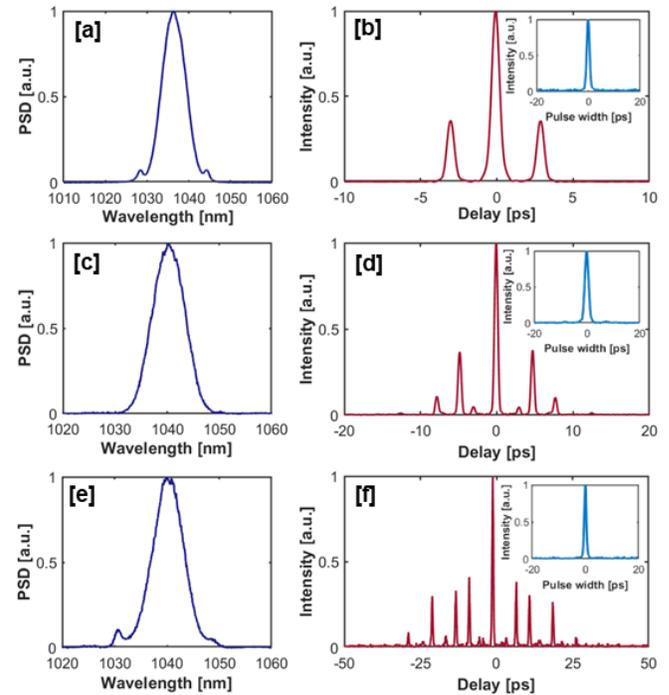

**Fig.2:** [a]: Spectral bandwidth at Port T with FWHM of ~8.2 nm for one division in PM fiber [b]: Corresponding AC trace of the pulse burst at Port M; the inset shows the autocorrelation of the recombined pulse at Port T with 0.8 ps duration. [c], [d]: Port T spectral bandwidth (8.3 nm FWHM) and AC trace of the pulse burst at Port M for two pulse divisions with additional 10 mm YVO$_4$-crystal, respectively. The AC trace of the recombined pulse at Port T (Inset [d]) has a FWHM of 0.83 ps. [e], [f]: Port T spectral bandwidth (8.2 nm FWHM) and Port M pulse burst autocorrelation for a third division with an additional 20 mm YVO$_4$-crystal. Inset [f]: AC trace of the recombined pulse at Port T with 0.8 ps FWHM.

For the DPO in Fig.1, the pulse divider consists of a 10 mm and a 20 mm long YVO$_4$-crystal in series in combination with the birefringent PM fiber that causes a total delay of ~2.8 ps. The delay caused by the YVO$_4$ is ~0.8 ps/mm at 1030 nm due to the large material birefringence of B=0.208. The fast axis sequential rotation angles in the DPO pulse divider with respect to the transmission axis of PBS1 for the two YVO$_4$-crystals and the PM fiber are 45°, 0° and 45°, respectively. Fig.1 [b] shows the resulting shape of $T(\Delta\varphi_{nl})$ for different values of phase-bias settings derived through an implementation of the YVO$_4$-crystals into the model of Ref. [10] with the corresponding Jones-matrices for the birefringent elements. A simulation of the pulse burst after the first 45°-rotated, 10 mm YVO$_4$-crystal can be seen in Fig.1 [c]. The phase-bias waveplate angles of θ=10° and ξ=25° ensure an energy splitting ratio of ε=0.8 with a positive slope of $T(\Delta\varphi_{nl})$ for small-signal values of $\Delta\varphi_{nl}$. Fig.1 [e] and [f] show the simulated divided pulses with two and three divisions after the subsequent, 0°-rotated 20 mm YVO$_4$-crystals and the PM-fiber segment at Port M, respectively.

In the fundamental mode-locked state of the oscillator the pulse division results of the 2.2m PM-fiber segment. Here, the laser generates soliton pulses with an output energy of 89 pJ measured at Port T with a QWP2 rotation angle that ensures an output coupling ratio of 30%. Fig.2 [a] shows the spectrum of the soliton with a FWHM of ∼8.2 nm centered at 1038 nm. Fig.2 [b] shows the autocorrelation (AC) trace of the pulse burst measured at Port M with a ∼2.8 ps delay for the orthogonal polarization modes after a single-pass through the PM-fiber. Simultaneously, the AC trace of a single recombined pulse can be measured at Port T with a FWHM of 0.8 ps. In this state, the laser is self-starting at a pump power of ∼300 mW, single-pulse operation can be achieved by reducing the pump power to ∼75 mW. The verification of stable single-pulse operation is based on multiple separate measurements. As a first step, the 150 ps wide-range autocorrelation together with a 0.02 nm high-resolution spectral measurement is used to ensure the absence of periodic modulations in the spectrum.

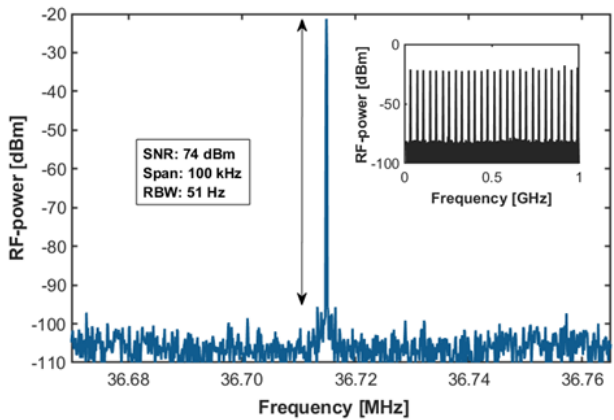

**Fig.3:** RF-spectrum of the DPO fundamental repetition-rate with 3 divisions at 36.7 MHz measured over a span of 100 kHz with 51Hz RBW. The SNR is ∼74 dBm. The inset shows the higher harmonics over a 1 GHz span (RBW: 100 kHz).

Further, the pulse train at Port T is detected with a fast InGaAs PIN photodetector (EOT-3000) and the radiofrequency (RF)-spectrum is measured with a spectrum analyzer (Keysight N9000A) over the full photodetector bandwidth of 2 GHz (100 kHz resolution) to verify the absence of amplitude-modulations of the higher harmonics. Fig.2 [c] shows the optical spectrum at Port T with a second pulse divider in the cavity. Here, the polarization modes are first separated in the 45°-rotated, 10mm YVO$_4$-crystal with a delay of ∼8 ps and then further divided in the 0°-rotated PM-fiber segment. The corresponding autocorrelation of the pulse burst measured at Port M is shown in Fig.2[d] together with the autocorrelation trace of the recombined pulse at Port T with a FWHM of 0.83 ps and a pulse energy of ∼190 pJ. Self-starting operation requires a pump power of ∼700 mW while the single-pulse threshold is at ∼160 mW. The higher pump power is required for an accumulation of sufficient small-signal $\Delta\varphi_{nl}$ despite the reduced peak power in the fiber segment due to the pulse division. For the third division, the 20mm YVO$_4$-crystal is implemented in the cavity with a 0°-rotation angle between the 45°-rotated 10 mm crystal and the 45°-rotated fiber, again with respect to the transmission axis of PBS1. The resulting spectral bandwidth (8.2 nm FWHM) together with the corresponding autocorrelations of the pulse burst at Port M and that of the recombined pulse at Port T (0.8 ps FWHM) can be seen in Fig.2 [e] and [f], respectively. For this state, the output pulse energy is increased by 6.5 dB to 400 pJ compared to the fundamental operation without YVO$_4$-crystals. Starting the laser in this state requires a pump power of up to 1.5 W with the single-pulse threshold at 400 mW. Further divisions were not possible in the experiment due to the lack of available pump power for an initiation of the mode-locked state.

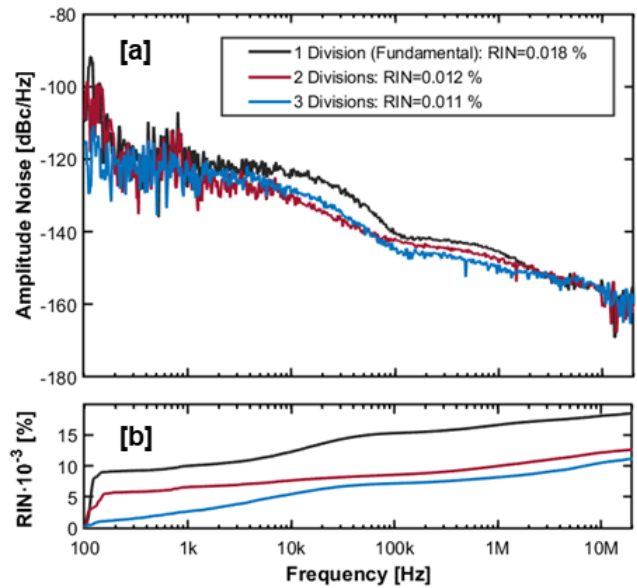

**Fig.4:** [a]: Frequency-resolved AM noise spectra measured at Port T with different pulse divisions in the cavity. [b]: Corresponding RIN integrated from 100 Hz to 20 MHz for the fundamental state, 2 divisions and 3 division with RIN values of 0.018%, 0.012% and 0.011%, respectively.

A significant aspect for the analysis of the DPO characteristics is the influence of each consecutive pulse division on the overall cavity stability and the noise performance. Fig. 3 shows the RF-spectrum of the fundamental frequency of 36.7 MHz measured at Port T with 3 pulse divisions in the DPO cavity corresponding to the working point shown in Fig.2[e] and [f]. The signal-to-noise ratio (SNR) is ∼74 dBm with a span of 100 kHz and a resolution bandwidth (RBW) of 51 Hz. The inset of Fig.3 shows the broadband RF-spectrum with a span of 1 GHz measured with a RBW of 100 kHz. Furthermore, the amplitude-noise (AM-noise) of the recombined pulse train emitted at Port T for different divided states of the oscillator is measured with the method described in Ref. [27]. Once the optical pulse train is converted to the electrical RF-domain with a fast photo-detector (EOT-3000), the 5th harmonic at 183.5 MHz of the received RF-signal is filtered with a tunable bandpass filter and amplified with a 10dB trans-impedance amplifier to an RF-power of -1 dBm. Subsequently, the AM-noise function of a signal-source analyzer (SSA) is used to measure the frequency-resolved noise spectral density. The experimental results for the AM-noise measurements at the recombined output Port T in the range from 100 Hz to 20 MHz can be seen in Fig.4 [a]. Here, the AM-noise spectra are shown for the fundamental mode-locked state with only the PM-fiber division (black) in comparison to the one measured with the second consecutive 10 mm $YVO_4$-division (red) and the third 20 mm $YVO_4$-division (blue). To minimize the influence of the pump source on the measured noise spectra, the laser is pumped with the same one of the two available laser diodes to maintain a stable single-pulse operation. As shown, in the high-frequency range from 100 kHz to ~3 MHz the AM-noise is significantly reduced (by up to 3 dB) with 2 pulse divisions in the cavity and by up to 6dB with 3 divisions, most likely as a consequence of the increased intracavity power as proposed in Ref. [18]. The AM-noise reduction is also present in the mid frequency-range from 10 kHz to 100 kHz by up to 9dB with the 10 mm crystal in the $YVO_4$-divider and up to 7 dB for the third division with the additional 20 mm crystal. Fig.4 [b] shows the corresponding relative intensity noise (RIN) integrated from 100 Hz to 20 MHz, excluding the influence of the technical noise sources <100Hz. For the fundamental mode-locked state, the RIN has a value of 0.018% that is consecutively reduced to 0.012% and 0.011% for the second and third division, respectively.

In conclusion, an all-PM fiber oscillator is investigated that uses periodic pulse division and recombination for the reduction of excessive roundtrip nonlinear phase shifts in the fiber segment. The implementation of the pulse division is enabled by the linear cavity structure that allows for a compensation of the linear phase shifts in the $YVO_4$-based pulse divider and the PM-fiber segment based on the application of a Faraday-mirror. In the experiment, the pulse energy at the output port can be increased by 6.5 dB from 89 pJ in the fundamental state to more than 400 pJ with 3 divisions. The stability of the mode-locked DPO for the consecutive pulse divisions is investigated through the RF-characteristics and the AM-noise of the output pulse train. The measured AM-noise is found to be reduced by up to 6dB in the high offset-frequency range from 100 kHz to 3 MHz. The improved pulse energy and noise performance that results from the implementation of the coherent pulse division and recombination is a major advantage of this cavity structure. The combination with established energy scaling methods based on dispersion-engineering and the application of LMA-fibers opens up new possibilities to extend the currently existing boundaries of reliable ultrafast all-PM fiber oscillators.

**Funding.** Deutsche Forschungsgemeinschaft (KA 908-1MUJO), European Research Council (609920).

**Disclosures.** The authors declare no conflicts of interest.